# From Content to Strategy: Understanding the Motivations, Processes, and Impacts of AI-Guided Communication

Chang Wan[1], Angel Hsing-Chi Hwang[2]

[1] Zhejiang University, Hangzhou, China
[2] University of Southern California, Los Angeles, CA, United States

E-mail:
chang.wan5@hotmail.com or chang.wan@zju.edu.cn
angel.hwang@usc.edu

## Abstract

Artificial intelligence-mediated communication (AI-MC) is conceptualized as applying AI to augment or generate message content (Hancock et al., 2020). However, advances in generative AI have expanded its use beyond generating content to guiding individuals' communication strategies, that is, AI-guided communication, yet theoretical and empirical understandings of this emerging use pattern and its consequences remains limited. To address this gap, this study conducted 26 in-depth interviews with individuals who have used AI to develop their communication strategies. Findings suggest participants strongly preferred using AI to analyze challenging scenarios in close relationships, because it fostered self-reflection, eased emotions, prevented conflict escalation, offered multiple perspectives, and provided a safe, nonjudgmental space for self-disclosure. Participants also stated that AI-guided communication enhanced their empathy and communication skills, though some voiced self-doubt and worried about losing their uniqueness. Views on long-term relational impact were mixed, depending on perceived usefulness of AI for resolving short-term interpersonal challenges.

Interpersonal communication is key to forming meaningful connections and close relationships, serving as a foundation for happiness, psychological health, and well-being (Lucas and Dyrenforth, 2006). With the rise of artificial intelligence (AI), many have applied it to interpersonal communication, such as moderating, refining, and even generating message content. This attracts scholarly attention and leads to establishing the body of work on AI-mediated communication (AI-MC, Hancock et al., 2020). Existing research has shown both potential and limitations of AI-MC in interpersonal communication settings (Fu et al., 2024). With growing capabilities of AI, users have also attempted alternative uses, including AI-guided communication.

In this paper, we refer to AI-guided communication (AI-GC) as applying AI to generate communication strategies and determine what to address in various interpersonal settings (Meng et al., 2025). Unlike AI-MC, which focuses on composing messages with AI assistance, users of AI-GC prioritize applying AI-generated output as guidance to tackle various challenges in interpersonal communication, while retaining their agency and autonomy to articulate their own messages. Recent research reveals that users might prefer AI-GC over AI-MC, as the former enables them to maintain greater independence during communication (Meng et al., 2025). Still, given the recency of the body of work, users' adoption and the potential impact of AI-GC remain largely underexplored.

The current work addresses this literature gap by identifying individuals' motivations, experiences, and processes to apply AI-GC. Moreover, we examine the downstream influences on interpersonal relationships as well as users' perception of the self and others, considering AI assistance plays a more prominent role in their social life. We interrogate these inquiries through 26 semi-structured interviews with individuals who have applied AI-GC in a wide range of interpersonal communication settings. Findings reveal that participants valued AI-GC as a supportive tool in close relationships, appreciating its ability to foster self-reflection, regulate emotions, prevent conflicts, and provide nonjudgmental spaces for self-disclosure. It also helped them enhance empathy and communication skills. Participants anticipated mixed impact of AI-GC on interpersonal relationships in the long run. While some participants viewed AI guidance as useful mainly for short-term conversations, others believed it enhanced relationship quality by promoting clearer expression, greater emotional understanding, and more mindful communication.

This study advances research on communication scholarship in several ways. First, it theorizes the emerging phenomenon of AI-GC by extending AI from a content-generation tool to a strategic consultant and reflective interlocutor that users draw on to interpret situations, generate strategies, regulate emotions, and rehearse interpersonal interactions while retaining agency over final decisions and communication execution. It also extends AI-MC theorizing beyond cognitive and message-level outcomes by foregrounding emotion regulation, like calming users, reducing anger, and preparing them to engage others more rationally, as a central rather than peripheral function of AI-GC. Further, it challenges assumptions that AI involvement signals reduced effort and undermines intimacy, showing that perceptions of AI use in relationships depend on how AI is used, with reflective and strategic support often interpreted as a sign of care rather than reduced effort. Finally, the findings reveal a self-limiting dynamic in which successful AI-GC may gradually reduce users' dependence on AI as they internalize transferable communicative skills, thereby extending conventional assumptions about technology adoption and sustained use.

## Literature Review

### AI-Guided Communication

In face-to-face settings, the content, style, and strategies of communication are fully crafted and controlled by humans (Hancock et al., 2020). However, advances in AI and large language models (LLMs) have opened new possibilities for refining communication content in real time, giving rise to AI-MC. Hancock and colleagues (2021) define AI-MC as "mediated communication between people in which a computational agent operates on behalf of a communicator by modifying, augmenting, or generating messages to accomplish communication or interpersonal goals" (p. 102). From this perspective, AI functions as a medium that not only supports interpersonal communication but also modifies, refines, and even generates messages.

While the use of AI-MC might facilitate and expedite message production, prior research demonstrates that people often perceive AI-generated messages as lacking emotional connection and social attraction compared to human-written content (Fu et al., 2024; Meng et al., 2023; Mieczkowski et al., 2021). Furthermore, message recipients could portray individuals who apply AI-MC in interpersonal settings negatively (Hohenstein et al., 2023; Mieczkowski et al., 2021). Consequently, individuals may be reluctant to rely on AI for composing messages.

Alternatively, individuals may be more accepting of using AI for guidance, ideas, and strategies to navigate interpersonal challenges (Fu et al., 2024; Meng et al., 2025). For instance, users might consult AI on whether to apologize or stand by their opinion after an argument, rather than directly composing a response with AI assistance. Meng and colleagues (2025) found that growing numbers of users actively seek AI suggestions to steer how they handle interpersonal scenarios, without necessarily involving AI in generating or modifying content. Fu et al. (2024) similarly found that AI tools can help users brainstorm new ways to respond to messages.

Meng and her colleagues (2025) introduce AI-GC to describe practices where humans actively ask AI for communication strategies. This form of communication differs from traditional computer-mediated communication (CMC), which treats computers merely as a medium (Herring, 2002); from existing conceptions of AI-MC, where AI is directly involved in creating specific message content; and from human–machine communication (HMC) (Hancock et al., 2020), where AI itself acts as a communication agent (Guzman and Lewis, 2020). Based on previous literature, we identify three defining characteristics of AI-guided interpersonal communication. First, both humans and AI are actively involved in the communication process (Meng et al., 2025). Second, AI's role lies not in providing content-level suggestions but in supporting broader communication strategies, such as initiating conversations, addressing conflicts, and rehearsing interactions (Fu et al., 2024). Third, humans retain agency over when to engage AI as a consultant and how to integrate its suggestions (Meng et al., 2025).

Aside from its formal definition and evidence suggesting that AI-GC is a more appealing and favorable form of support than AI-generated or AI-modified content, much of the topic remains underexplored. It remains largely unclear what drives individuals to seek AI assistance for communication strategies, why people perceive it as effective, and how it facilitates interpersonal communication. To address this literature gap, the present study focuses on investigating how individuals draw on AI to guide their interpersonal communication strategies. Extending the above-mentioned literature, we propose the following research questions:

- **RQ1:** What factors drive individuals to seek advice from AI for interpersonal communication?
- **RQ2:** What are the major functions of AI-GC when people use generative AI to seek advice on communication strategies in interpersonal communication?

**Potential of AI-Guided Interpersonal Relationship Management**

While previous literature has noted the potential of AI-MC to enhance interpersonal communication, little research has examined its subsequent impacts, such as on relationships and users' self-identity. On the one hand, improved communication is often assumed to foster stronger relationships (Pang et al., 2018; Szkody and McKinney, 2021). In this sense, AI-GC may provide an avenue for individuals to better manage their interpersonal relationships. On the other hand, unlike traditional AI-MC, which involves direct content generation, AI-GC enables users to retain agency by composing messages themselves while selectively drawing on AI's suggestions (Meng et al., 2025). This raises important questions about how individuals perceive the impact of AI-GC on the self.

Interpersonal relationships entail formal and informal ones (Fu et al., 2024). Formal relationships, such as workplace relationships, are often established in professional and formal environments and shaped by workplace norms, roles, and responsibilities, and characterized by limited personal sharing. Informal relationships, such as friendships, romantic relationships, and family relationships, indicate relationships with high levels of casual language, emotional closeness, and personal sharing (Fu et al., 2024). Interpersonal communication within informal relationships serves as a primary channel for expressing emotions, disclosing oneself, seeking social support, and ultimately enhancing well-being (Liu et al., 2024). This study therefore focuses on how AI helps manage informal relationships.

Meng et al. (2025) found that in social support contexts, users perceived AI-guided suggestions as more helpful and authentic than AI-generated or AI-modified content, because they retained agency and their own voice. However, Fu et al. (2024) have found that users tend to hold mixed feelings in informal relationships. Some participants felt that AI-MC tools helped them communicate more thoughtfully, while others viewed such practices as inauthentic. Prior studies do not explain why views diverge. We therefore propose the following research question:

- **RQ3:** How do individuals perceive AI-guided interpersonal communication, and how do these perceptions influence their adoption of generative AI?

AI has long been associated with a positivity bias, often making language appear warmer (Hwang and Won, 2021). Mieczkowski et al. (2021) found that AI-MC tools make users' messages more positive, and Fu et al. (2024) found that users often self-reflect, adjust their writing style, and become kinder when aided by AI. Yet it remains unknown whether these shifts in language influence users' self-concepts.

Additionally, Hancock and his colleagues (2021) raised concern that AI use may signal less effort in a relationship, which may damage intimate relationships. Supporting this, Liu et al. (2024) found through an experiment that when AI intervenes in interpersonal communication, users are seen as exerting less effort by turning to AI to modify or generate content, which can lower relationship satisfaction.

According to the discussion above, the following research question is proposed:

- **RQ4:** How do AI-generated suggestions shape individuals' self-concept, communication practices, and interpersonal relationships?

## Human versus AI Guidance in Interpersonal Communication

AI-GC can provide both emotional and informational function (Meng et al., 2025). For informational function, humans draw on personal experience, which is rich in detail but limited in scope and shaped by their own knowledge and experiences (Boyacı et al., 2024), whereas generative AI draws on broad sources and data to offer comprehensive, multi-perspective, and sometimes novel suggestions (Fu et al., 2024; Meng et al., 2025). Additionally, when users seek help from people around them, they often receive empathy and agreement (Finkel et al., 2017; Reid et al., 2021). AI, however, is generally perceived as more neutral and objective, offering multiple perspectives that can help individuals better understand situations and manage their interpersonal communication (Cheong et al., 2025; Fu et al., 2024). However, interpersonal communication management is inherently personalized and contextualized (Mou and Xu, 2017). People close to the user often have richer knowledge of their personality, past experiences, and living environment, which allows them to provide more tailored strategies (Brandtzaeg et al., 2022). In contrast, AI, relying on large datasets, tends toward generalized advice that can feel mismatched or impractical (Fu et al., 2024). Another distinction lies in availability. Human support is limited by time and cognitive capacity, while AI is available 24/7 and consistently patient (Brandtzaeg et al., 2022).

Regarding emotional function, Meng et al. (2023) found that human support providers perform better than chatbots, facilitating cognitive appraisal and reducing distress more effectively. However, Fu et al. (2024) reported that using AI-MC tools can also enhance users' confidence in interpersonal communication and provide positive reinforcement that makes them feel comforted. Additionally, AI is increasingly perceived as capable of providing emotional support (Meng et al., 2025; Yin et al., 2024). Liu et al. (2024) likewise found no significant difference in relationship satisfaction between advice from AI and from humans. We therefore propose the following research question.

- **RQ5:** How do people compare their experiences of seeking communication and relationship advice from AI versus from humans?

## Method

The current work aims to investigate this emerging phenomenon where users increasingly apply AI to provide strategies for interpersonal communication and relationship management. To address our research questions, we conducted semi-structured interviews with 26 users who have sought AI for communication-related advice. As Jiang et al. (2021) and Guzman (2023) suggested, interviews are effective for generating profound insights and enhancing our understanding of AI-related phenomena; it is also a suitable method given the exploratory nature of the current research topic (Brandtzaeg et al., 2022).

## Sample and Recruitment

We recruited active users of generative AI between the ages 19 and 30, who match the demographic profile of the majority of users on generative AI such as ChatGPT. Snowball sampling and purposive sampling were adopted to recruit participants. Initially, we reached out to potential participants through postings that we are looking for users who have asked generative AI for suggestions on interpersonal communication in social media platforms such as Rednote, Reddit, and Facebook. Since having AI assist in interpersonal communication management is a relatively new phenomenon, even a small sample can yield comprehensive and valid insights (Brandtzaeg et al., 2022). See Table 1 for details of the participant profile. We stopped recruitment once we reached empirical saturation. Each participant was rewarded with 30 RMB.

We interviewed 16 females and 10 males. Most of them are Chinese ($n = 21$), and participants from South Africa and Kenya who live in the US ($n = 5$). Interviews with Chinese participants were conducted in Chinese and the others in English, and each was audio-recorded for verbatim transcription.

**Table 1** *Demographics of Participants*

| ID | Pseudonym | Age | Gender | Ethnicity | Occupation |
|---|---|---|---|---|---|
| 1 | Tina | 26 | Female | Chinese | National Serviceman |
| 2 | Phoebe | 26 | Female | Chinese | National Serviceman |
| 3 | Sally | 26 | Female | Chinese | National Serviceman |
| 4 | Wendy | 26 | Female | Chinese | Bank Worker |
| 5 | Amy | 26 | Female | Chinese | Unemployed |
| 6 | Kelly | 26 | Female | Chinese | Channel Manager |
| 7 | Linda | 26 | Female | Chinese | Headhunter |
| 8 | Jessica | 22 | Female | Chinese | Student |
| 9 | Coco | 25 | Female | Chinese | Teacher |
| 10 | Lisa | 26 | Female | Chinese | Chemical Plant Employee |
| 11 | Cindy | 24 | Female | Chinese | Student |
| 12 | Carol | 25 | Female | Chinese | Sales |
| 13 | Leon | 26 | Male | Chinese | Operations Specialist |
| 14 | Jimmy | 27 | Male | Chinese | Automotive Technician |
| 15 | Carina | 19 | Female | Chinese | Student |
| 16 | Maggie | 30 | Female | Chinese | Teacher |
| 17 | Jack | 24 | Male | Chinese | Host |
| 18 | Richard | 21 | Male | Chinese | Student |
| 19 | John | 21 | Male | Chinese | Student |
| 20 | Thomas | 22 | Male | Chinese | Student |
| 21 | Charles | 19 | Male | Chinese | Student |
| 22 | William | 25 | Male | American | Designer |
| 23 | Michael | 27 | Male | Ugandas | Student |
| 24 | James | 24 | Male | South African | Computer Scientist |
| 25 | Alex | 25 | Male | Yemeni | Sales |
| 26 | Daisy | 26 | Female | African | Accountant |

### Procedure and Interviews

Interviews took place between June and August 2025 via Tencent Meeting (Chinese participants) and Zoom (others), lasting roughly 60 to 90 minutes. All participants gave informed consent for the interview and audio recording.

Guided by a semi-structured interview protocol (see Appendix), participants described the interpersonal scenarios in which they turned to AI and their motivations for doing so. They then shared successful and unsuccessful examples of applying AI suggestions and their criteria for accepting or rejecting them and assessed the impact of AI on themselves and their relationships. Follow-up questions compared seeking help from people versus AI, and asked about lessons from prior experience and intentions for future use. Participants were then debriefed and thanked.

### Analysis

First, during the open coding stage, we familiarized ourselves with the transcripts and generated initial codes for meaningful and recurring statements (e.g., availability of AI), yielding 84 codes. Second, we reviewed all open codes and grouped them into higher-level, second-order codes or themes. This categorization was guided by the above-mentioned AI-MC, HMC work from our literature review section, and our research questions. Third, we discussed and refined these second-order themes by integrating overlapping themes, excluding rarely mentioned ones, and further theorizing them to address our research questions. Data analysis reached saturation when the themes in our conceptual framework were adequately represented in the data (Saunders, 2018).

## Results

### Understanding Motivations via Comparison with Humans (RQ1 & RQ5)

We initially proposed RQ1 and RQ5 to examine individuals' motivations for seeking AI support and the differences between AI versus human support. However, during our interviews, participants consistently framed their motivations by comparing AI with asking other people for help. Therefore, we address these two questions together.

**Availability and safety.** Previous literature has extensively examined the advantages of AI in its availability and feeling safe for people who are willing to communicate with it and even develop friendships with AI (Brandtzaeg et al., 2022). Additionally, most participants said communicating with AI gave them a sense of security, especially because it protects their privacy, letting them express themselves freely (Fu et al., 2024). This study provides support to existing literature.

**Feeling stressless.** Consistent with prior work showing users feel judgment-free with AI (Le et al., 2025; Yam et al., 2023), our participants felt less pressure when communicating with AI. Beyond that, our findings further suggest that AI reduces communication anxiety by helping users overcome shyness. For example, Jimmy (P14) shared, "When I ask other people for help, I feel embarrassed, like I owe them a favor. But you don't have such feelings when asking AI for help." Some participants noted that AI relieved them of the worry that asking too many questions might annoy others. Leon (P13) added, "Sometimes when you share too many emotions or ask too many questions, people might become a little impatient, or if you repeat something a few times, they might point it out. But AI would never get impatient."

**Neutral perspective.** Although prior work suggests users seek AI for objective recommendations in decision-making (von Eschenbach, 2021), our participants expressed a more specific and somewhat different set of motivations when explaining why they sought AI-generated interpersonal communication advice. Most participants shared that turning to friends or family for relationship advice often yields unconditional support that discourages questioning one's own judgment. In contrast, AI is perceived as a neutral and rational third party, helping users analyze situations more clearly. For instance, Kelly (P6) explained, "With female friends, they'll almost always take your side and jump to the conclusion that 'the guy's just not good.' However, I still wanted an objective third-party perspective to help me analyze it." Carina (P15) added,

> Sometimes I wonder if I did something wrong, but what I really want is an objective answer. And honestly, only AI can give me that (answer). When it comes to friends or family, they'll usually just back you unconditionally, like "if you want to do it, then do it."

**Multi-dimensional perspectives**. Notably, participants highlighted that since people often rely on their own experiences to provide suggestions, such advice is naturally limited because one could only speak from their own life experience. In contrast, AI can draw from large datasets to offer more comprehensive strategies. Loen (P13) stated, "Sometimes it just gives a pretty complete take, and I feel like people don't really have that kind of experience, so they're not quite there in terms of expertise yet." Sally (P3) explained,

> I think AI looks at things from a comprehensive perspective. If I'm thinking on my own, I might only come up with a few points—it's limited. But AI can look at it from multiple angles and help me analyze it.

**Turning to AI as a last resort.** Many participants said turning to AI signals they have no other way to solve an interpersonal problem. Lisa (P10) stated: "I only turn to AI when I really have no other options. If I already have some ideas, I wouldn't ask it." Phoebe (P2) added: "I wasn't very satisfied with the answers given by other people, so I turned to AI."

**Functions and Contexts of Applying AI-Guided Communication (RQ2)**

RQ2 examines *how* users applied AI suggestions and guidance to strategize interpersonal relationship management. Consistent with prior research, users engaged AI to analyze situations, detect subtle signals, and generate new ideas for responding (Fu et al., 2024; Meng et al., 2025). Notably, our findings add three insights. First, preferences for traditional AI-MC versus AI-GC varied across relational contexts, shaped by perceived social distance and the need for self-disclosure. Specifically, workplace or tutor-student relationships were seen as distant and low in self-disclosure, whereas close relationships invited openness and psychological closeness.

In formal settings, such as workplaces, participants prefer AI-MC tools that directly generate or modify content. Here, participants focused on generating polite content and avoiding errors in messages to appear "perfect" and professional (Fu et al., 2024). As AI-generated content was sufficiently polite and formal, relying on AI-MC helped reduce their cognitive burden to craft error-free messages. In contrast, in informal or close relationships, participants show a stronger preference for AI-GC. In these contexts, participants felt less pressure to be flawless and were more willing to reveal their authentic voices. Here, they focused on seeking AI suggestions to address interpersonal hurdles they were otherwise unsure how to handle, making traditional AI-MC functions less relevant.

Second, participants described a sequential pattern in how they engaged AI. They typically began by using AI to analyze complex interpersonal challenges they faced, including gauging other people's thoughts, detecting subtle cues, and facilitating self-reflection. The analysis part is crucial in AI-GC, as one of the participants (Sally, P3) described it as "The analysis suddenly made me aware of my blind spots, revealing deeper issues in the relationships that I likely would not have recognized on my own without the AI's insights." They then prompted AI to suggest strategies, organize their thoughts, or rehearse upcoming conversations.

Third, generating ideas or strategies was the most common function for informal relationships. Participants generally disliked having AI polish or generate actual wording; instead, they preferred to formulate their own expressions while drawing on AI's ideas. Participants consistently noted that AI-generated content often feels artificial and "machine-like," making it unsuitable for direct use in close relationships such as friendships or romantic partnerships. As Carol (P12) explained, "I usually take the general direction from AI, but I often feel the exact wording is not precise enough, so I adjust it." Jack (P17) added, "What the AI gives me isn't so much new wording—I could have said those words myself—but rather a perspective I hadn't considered before."

We found that people seek different types of assistance depending on the context. In uncertain situations such as quarrels, participants expected AI not only to provide suggestions but also to analyze the situation and offer emotional support. In low-uncertainty situations such as planning a surprise, they mainly wanted AI to generate ideas directly, without analysis or emotional support. Additionally, we also observed cultural variations. Non-Chinese participants more often used AI for positive purposes, such as ideas to surprise partners, whereas most Chinese participants turned to AI mainly to address conflicts. The exception was those in the dating stage, who often sought AI's guidance on how to communicate with potential partners and develop their relationships further, a context that also involves considerable uncertainty, where analysis and situational insights remain important.

One of the Chinese participants (Maggie, P16) who used ChatGPT reflected on cultural differences in communication, stating: "I think Eastern people are often more reserved or indirect. In contrast, ChatGPT, rooted in a Western mindset, tends to be straightforward and encourages me to say what I want. Actually, this directness actually helps reduce misunderstanding."

**Attitudes toward Using AI during Interpersonal Communication (RQ3)**

Participants exhibited polarized attitudes toward others using AI to manage interpersonal communication. A small number viewed it as insincere and distancing, even though all participants themselves have used AI for communication suggestions. Phoebe (P2) explained, "Because I still hope people get to know me through my own expressions, not through AI. Otherwise, it feels to me like they are taking a shortcut in the process of building a relationship with me." Maggie (P16) added, "Using AI to assist communication feels like a perfunctory act."

In contrast, most participants regarded the use of AI as a sign of genuine effort to develop or sustain a relationship. Cindy (P11) put it, "I feel that his use of AI shows he is putting a great deal of effort into understanding and maintaining our relationship. To me, this suggests that he wants to continue with me." Amy (P5) noted, "I think I would feel better, because it would show me that the other person is taking our conversation seriously."

Kelly (P6) shared a particularly emotional account of discovering that her mother had consulted AI about how to communicate with her:

> It felt really warm—I almost wanted to cry right then. My mom was working so hard to get closer to me, and when she chatted with AI, it really felt like talking to a real person—so polite and genuine.

**Impacts on Self (RQ4)**

Most participants shared that AI-GC had positive effects on themselves, including teaching them communication skills, giving them courage and confidence to express themselves, and fostering peace of mind.

**Peace of mind through situation analysis**. Before offering suggestions, AI often demonstrated empathy and helped participants analyze the situation. By clarifying what happened, participants could calm themselves and communicate more peacefully, rather than being driven by anger into hurtful words or irrational decisions. Amy (P5) explained, "It feels like what AI really does is help you see things more clearly, and that leads you into a process of self-reconciliation. And that process of reconciling with yourself makes it easier to communicate with others." Similarly, Jessica (P8) reflected, "It helped me let go of conflicts. Instead of overthinking and making myself upset, I could move on, feel lighter, and then talk to others more easily. It really made communication smoother."

**Enhancing intimacy and empathy.** By analyzing others' subtle thoughts, AI helped participants become more empathetic and more willing to express themselves. Coco (P9) recounted:

> While living away from home, I was once scammed and lost all my money. When I told my parents, their reaction felt blaming and made me even more upset. After sharing the experience with AI, it suggested that their response likely stemmed from concern rather than blame. This helped me reflect on my own emotions and later explain them to my parents. They were understanding, which eased the tension between us. Since then, I have become more willing to share more with my parents.

In this sense, AI helps users better understand others' thoughts and clarify misunderstandings that may arise from language.

**Impact on self-identity.** As participants increasingly used AI for interpersonal challenges, they raised concerns about its implications for self-identity. The main worry was over-reliance, one of the most common issues discussed in the literature (Fu et al., 2024). Our findings further confirmed that, even in the context of interpersonal communication, people still worry about becoming overly dependent on AI, which may weaken their ability to think independently. Maggie (P16) even argued that they sometimes felt as if their partner was "dating the AI instead of themselves."

***Self-doubt and loss of uniqueness.*** A few participants also frequently mentioned that they started doubting themselves and questioning their identity of self. Coco (P9) shared,

> Sometimes I feel like I'm not really good at communicating, but with AI I sound like I am—is that really me? Then I wonder, if I stop using it, will I still be like that? And if people see the real me, will they accept it?

Others expressed concerns about losing their personality and uniqueness: "AI isn't just for me, other people can use it too, and they may use it as I do. If everyone's doing things in a similar way, then what new way do I have to solve these problems?"

***Internalization and retained agency.*** Regarding the positive effects, some participants felt that they could internalize and absorb these strategies and skills, believing that this knowledge became their own and helped them become better selves. They emphasized that they still retained agency. Maggie (P16) even summarized AI's response framework: "First, it will show empathy and give some reasons for why it understands you, and finally, it provides its suggestions." She applied it by first showing empathy, acknowledging the other's good intentions, then politely sharing her own thoughts. Additionally, Jimmy (P14) also had similar feelings, "AI taught me new ways to communicate, and I've absorbed those lessons. Now I can handle similar situations on my own."

## Impact on Relationships (RQ4)

**Limited effects but continued use.** Interestingly, about half of the participants reported that AI had no impact or only limited effects on their interpersonal relationships. For example, Lisa (P10) noted, "I don't think it has made my relationship better, but at least it hasn't made it worse." People may still rely on their own agency to determine how the relationship will develop, as Leon (P13) explained, "When certain problems arise, AI can sometimes provide useful assistance. However, the future of a relationship ultimately depends on the individuals involved, and AI cannot determine its direction."

Despite perceiving limited effects, most still wanted to use AI, partly because it assists instant, short-term conversations and helps sustain the status quo, preventing relationships from deteriorating. Carina (P15) noted, "Even if it doesn't change the relationship much, it still helps like making conversations better and making me less harsh. That's why I'll keep using it." Others continued for wanting emotional support. As Wendy (P4) explained, "I don't think it really made my relationship better. But it did help me with my emotions."

**Positive impacts but reduced use.** Half of the people would say that AI really helps them manage interpersonal relationships. For example, Jimmy (P14) says AI is like his tutor when approaching his girlfriend. Jimmy said,

> I honestly feel like it's thanks to AI that I'm even dating my girlfriend now. Before, when I was in relationships, I couldn't really pick up on hints or subtle signals. So, I kept asking AI what my next move should be.

Other people noted AI helped them refine their communication style, express themselves more effectively, and navigate challenging situations. “The best part is it really improved my relationships. I deal with things differently now and avoid a lot of conflicts.”

Since AI can help people analyze situations during conflicts, it often enables users to better understand others’ perspectives, thereby fostering empathy and closeness. Several participants noted that in conflicts with parents or partners, AI offered interpretations that helped them see the situation differently. Richard (P18), a blind participant, shared that on the day his grandfather passed away he wanted to attend the funeral, but his parents refused. He felt heartbroken, believing they saw him as a burden. However, AI offered another perspective. His parents knew he was sensitive and caring and feared the grief would overwhelm him, and worried about his safety in a crowd. This revealed emotions Richard had not considered, helped him understand his parents’ intentions, made him more willing to communicate, and ultimately brought him closer to his family.

Notably, participants who found AI helpful in their relationships also tended to use it less over time. First, they explained that having learned from AI, they could communicate effectively on their own. Second, some noted that because AI often gives positive feedback, they risked over-reliance and so became more attentive to self-control.

## Discussion

Existing AI-MC research has focused largely on how AI modifies, augments, and generates messages (Hancock et al., 2020). These functions are especially useful in formal settings, such as workplace or advisor-student interactions, where communication is professional and relational distance is expected. In such contexts, individuals value clarity, politeness, and accuracy, and AI tools help them produce well-structured and error-free messages. However, this message-centered approach is less effective in informal and close relationships (Fu et al., 2024). In personal relationships with friends, romantic partners, or family members, polished wording is not the primary concern. Instead, the goals of communication shift toward maintaining intimacy, managing conflict, and building mutual understanding. In these contexts, AI-generated or overly refined messages may be perceived as artificial or emotionally exaggerated (Fu et al., 2024). Rather than using AI to automatically generate responses, individuals often seek support that helps them think through their feelings and responses. Meng et al. (2025) describe this emerging approach as AI-GC, in which AI supports reflection and offers communication strategies rather than producing ready-made messages.

By shifting attention from message production to strategic support, this study synthesizes a conceptual framework that clarifies the motivations behind AI-GC and outlines its core process and its effects on users and relationships (See Figure 1). This framework offers a comprehensive understanding of how individuals engage with AI not simply to refine messages, but to navigate interpersonal challenges. In doing so, the study moves AI-MC scholarship beyond its focus on message generation and optimization and provides a clearer theoretical account of how AI can support reflection, strengthen relational skills, and contribute to interpersonal growth and relationship maintenance.

**Figure 1** *A Conceptual Framework of AI-Guided Communication: Motivations, Processes, and Impacts*

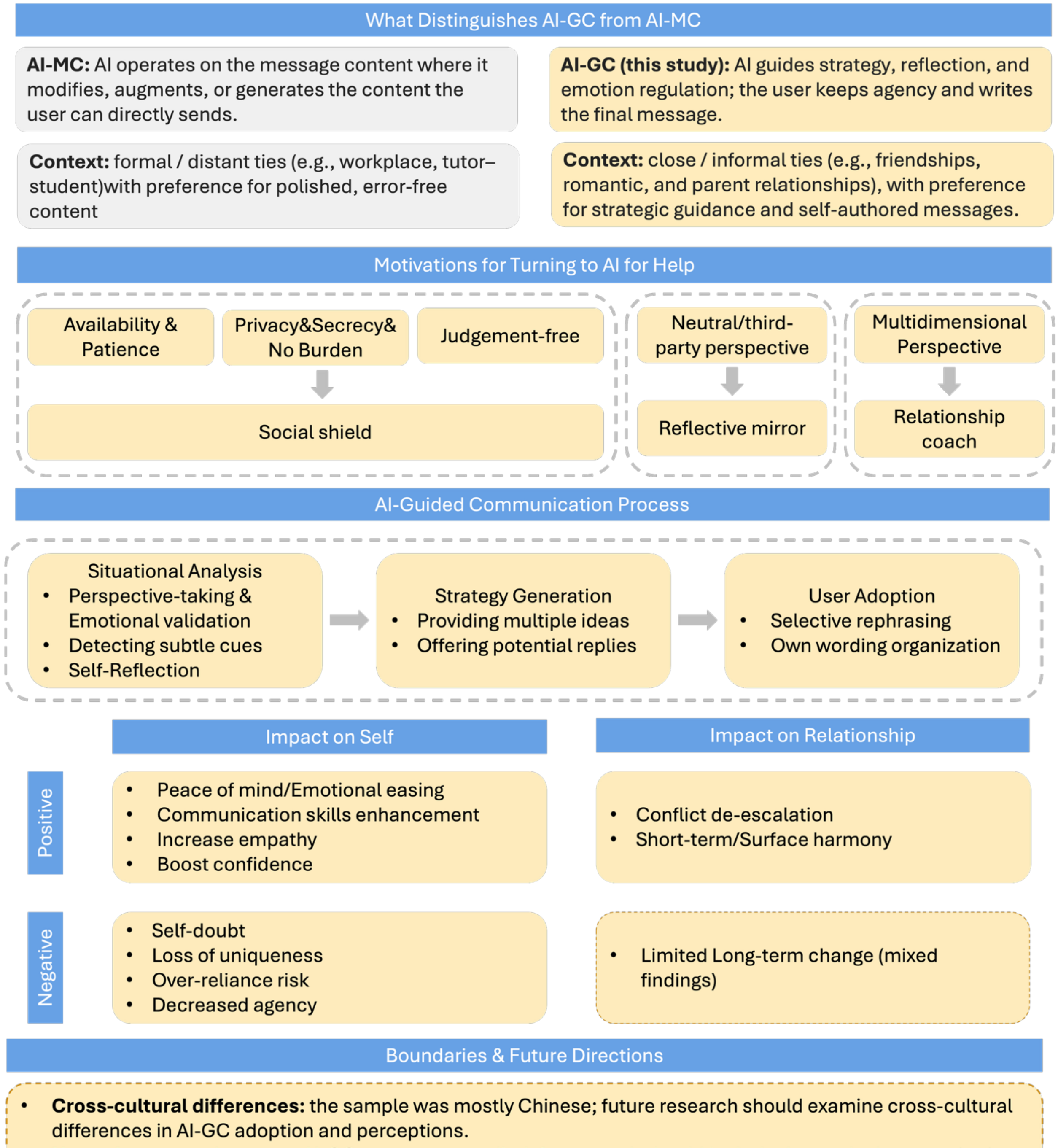

**Motivations and Perceptions of AI-Guided Communication**

Prior research has identified several common motivations for using AI-MC tools, including their constant availability and the privacy they afford (Brandtzaeg et al., 2022; Fu et al., 2024; Hancock et al., 2020). Beyond these, the present study uncovers additional motivations specific to AI-GC in interpersonal contexts. Participants described AI as offering a "stress-free" way to seek support. Participants valued being able to discuss sensitive or personal issues without fear of embarrassment or judgment, which helped them overcome reluctance to open up. Unlike seeking advice from another person, turning to AI also eliminated concerns about burdening others with their personal problems or exposing their vulnerable moments. AI, as participants noted, is "endlessly patient." Beyond emotional comfort, participants appreciated AI's capacity to offer broader and more balanced perspectives. Whereas close others often reinforce people's existing viewpoints, AI functions more like a neutral third party, helping participants step back and assess their situations more objectively. In addition, its access to extensive information enabled it to suggest options that might not occur within one's social experience. For some, AI served as a last resort when other forms of support were unavailable.

Our findings also challenge prior assumption that relying on AI signals indifference or a lack of genuine engagement. In contrast, most participants in this study viewed AI use as evidence of effort and commitment to the relationship. Drawing on their own experience, they noted that working with AI to develop a contextually appropriate response requires time, reflection, and cognitive effort. Rather than a shortcut, they interpreted this process as a meaningful investment, making AI-GC appear deliberate and sincere rather than perfunctory.

**The Process of AI-Guided Communication**

As AI-GC is still an emerging concept, little is known about how users actually engage with AI to develop communication strategies and put them into practice. This study clarifies this process by identifying its key stages and how they unfold in practice.

Users typically engage AI through a sequence of steps oriented toward managing interpersonal relationships. The process often begins with asking AI to analyze the situation, helping users understand what has happened and consider how the other person might be thinking or feeling. This interpretive step frequently has a calming effect, giving users the space to reflect on their own behavior and acknowledge problems. AI also provides emotional support during this stage, and together, contextual analysis and emotional support appear to prepare users to engage more openly with the advice that follows.

Theoretically, this suggests that AI-GC is not simply a strategy provider but also provides emotional functions. By regulating users' emotional states before they engage with the situation, it enables them to approach others with greater calm and openness, thereby reducing the risk of escalation and creating more favorable conditions for subsequent communication. This emotional regulation dimension has been largely overlooked in existing research on AI-MC, whereas our study brings it to the forefront and demonstrates its central role in shaping how users interpret and enact AI guidance.

After clarifying the situation, AI typically offered practical strategies for moving forward. Participants valued these suggestions for their breadth and perspective, noting that AI could draw on knowledge beyond their immediate social experience. Importantly, unlike traditional AI-MC tools that focus on drafting messages, AI-GC preserves user agency. Rather than copying AI-generated text, users learned from and internalized AI's suggestions, adapting them and expressing the ideas in their own words. This process allowed them to avoid machine-like language while retaining ownership of their communication.

### Impacts of AI-Guided Communication on Individuals, Relationships, and Continued Use

A notable contribution of this study lies in examining the effects of AI-GC on both individuals and their relationships. At the individual level, most participants did not perceive AI use as threatening their self-identity. Instead, they described it as helping them become more thoughtful and effective communicators. Participants characterized AI's responses as measured and constructive. For example, in conflict situations, AI often encouraged users to acknowledge the other person's perspective, even an opposing one, before stating their own. By highlighting possible good intentions behind others' actions, AI fostered perspective-taking, helped participants interpret situations more charitably, feel emotionally supported, and approach conversations with greater calm. However, a small number of participants expressed concerns about overreliance on AI, worrying that they might become unable to resolve interpersonal issues independently without AI.

At the relational level, perceptions were more mixed. Participants agreed that AI tended to recommend conflict-avoidant or de-escalatory strategies. For some, this style strengthened relationships by facilitating smoother communication and more constructive problem-solving. However, others viewed AI's influence as limited. They argued that while AI could assist with specific conversations, the long-term quality of a relationship ultimately depends on individual agency and mutual effort that is beyond AI's control. A few participants went further, suggesting that AI's preference for avoiding conflict might be counterproductive, concealing deeper relational problems rather than resolving them.

Participants' intentions to continue using AI also varied in a pattern with theoretical implications for our understanding of user behavior. Those who felt AI had improved their relationships were surprisingly more inclined to reduce their use over time. Some worried about becoming too dependent on AI in ways that might undermine both their own communication skills and the authenticity of their relationships. Others described their engagement with AI as a learning process, having internalized useful strategies, they now felt equipped to handle similar situations without assistance. In contrast, participants who saw AI's relational impact as limited were more likely to continue using it. For them, AI's value lay not in transforming the relationship, but in the reliable, ongoing support it provided in processing situations and managing emotions. This suggests a self-limiting dynamic in AI-GC, whereby perceived effectiveness reduces continued reliance, challenging conventional technology adoption models that treat sustained use as the primary marker of value.

### Implications

First, this study advances communication scholarship by reframing the locus of AI involvement in interpersonal communication. Whereas existing AI-MC research has predominantly conceptualized AI as a content-level agent that modifies, augments, or generates messages (Hancock et al., 2020; Mieczkowski et al., 2021), our findings indicate that users increasingly engage AI at the strategic level, drawing on it to interpret situations, generate communication approaches, and rehearse interactions while retaining agency over the final communicative decisions and execution. This shift from content production to strategic support warrants a broader theoretical lens, one that accommodates AI as a reflective interlocutor and strategic consultant rather than only as a writing assistant. The conceptual framework introduced here offers an initial step by linking users' motivations, the sequential process of AI-GC, and the downstream effects on the self and relationships.

Second, findings from this study expands the emotional and cognitive functions of adopting AI assistance in communication processes. Prior work has focused largely on cognitive and message-level outcomes such as perceived authenticity, social attraction, and language change (Hohenstein et al., 2023; Mieczkowski et al., 2021) as well as emotional support (Meng et al., 2025). Beyond these, AI assistance also helps users think through, clarify, and delineate both their feelings and the situational factors that they face in their interpersonal challenges. Additionally, our findings reveal that a primary function of AI-GC lies in emotional regulation: users described how AI's situational analysis calmed them, reduced anger, and prepared them to engage others more rationally. This affective scaffolding occurs prior to the interpersonal interaction itself, suggesting that emotion regulation is not merely a peripheral by-product but a central function through which AI-GC shapes interpersonal outcomes. Integrating emotion regulation into AI-MC models therefore offers a more comprehensive understanding of how users benefit from AI in relational settings.

Third, the study resurfaces prevailing assumptions about effort, authenticity, and AI use in relationships. Whereas earlier work has argued that AI involvement may signal reduced effort and undermine intimacy (Hancock et al., 2020; Liu et al., 2024), most participants in our study interpreted others' use of AI as a sign of care and investment, especially when it served reflection rather than outsourced expression. This finding suggests that perceptions of effort in AI-mediated contexts are contingent on the function AI serves content generator may be read as shortcut-taking, whereas reflective and strategic use may be read as deliberate effort.

Fourth, our findings surface a self-limiting dynamic that challenges conventional adoption models. Classic frameworks treat sustained or intensified use as the principal indicator of value (Venkatesh et al., 2003). Yet participants who perceived AI-GC as most beneficial were often those who reported decreasing reliance over time, having internalized AI's strategies as transferable communicative skills. Conversely, those who perceived more limited relational impact tended to continue using AI for emotional support and momentary problem-solving. This pattern suggests that AI-GC may serve as a scaffolding tool that helps users build transferable skills, with its success partly reflected in their growing independence from it.

**Limitations and Future Studies**

A major limitation of this study is that most participants were from China, although we did collect some data from foreign participants. Future research can extend this study by comparing cultural differences, particularly examining how Western participants' perceptions of AI-GC may differ from those of Chinese participants. Another promising direction would be to investigate the role of culturally rooted differences in chatbots, for example, how Asian users experience Western AI tools (Segerer, 2025). Second, because our research only recruited individuals who had already engaged in AI-GC, future studies should also include those who have not, or have not yet, adopted this practice to better capture diverse perspectives. Third, this study gathered insights only from users of AI-GC. Future work could expand by incorporating both message senders who rely on AI guidance and message recipients who interact with such AI-assisted communication to examine whether perceptions differ between the two groups.

## Appendix

### Interview Protocol

### I. Features, Motivations, and Process of AI-Guided Communication

1. Have you ever used a chatbot (e.g., ChatGPT) to seek advice for communicating with others? What kind of conversations or relationships (e.g., with friends, coworkers, family) have you used AI to support?
2. How often do you turn to AI for communication-related advice? What influences your frequency of use?
3. What motivates you to seek communication advice from AI tools?
4. In what ways do you typically use AI to support your communication? Which specific features or functions do you find most useful?
5. Have you ever applied AI-generated suggestions in real-life, face-to-face communication? Could you share an example where the AI's advice worked well?
6. Can you recall a vulnerable moment in one of your close relationships, when you turned to AI for advice? If an AI helped you navigate that moment, how did that affect your feelings of love, intimacy, or connection?
7. Please share one example where it didn't work as expected. What do you think caused the negative outcome?
8. What factors most influence whether you accept or reject AI's suggestions?

### II. Influences of AI-Guided Communication

9. Overall, how did you feel about using AI to support your communication? In what ways, if any, did it help or fail to help your conversations?
10. In what ways did it positively or negatively affect your relationships with others?
11. In your view, what are the main advantages and drawbacks of using AI to guide interpersonal communication?
12. After using AI suggestions, do you feel more capable of handling similar situations on your own in the future, or would you still prefer to rely on AI for guidance? Why?
13. How do you feel if others use AI tools to assist their communication with you? Why?
14. Do you feel emotionally supported by AI when preparing for communication? If so, in what ways?
15. Looking ahead, how do you see yourself using AI to support your future relationships and communication practices? Which specific functions would you use, and in what kinds of situations?
16. Conversely, in what situations would you choose not to use AI, and why?

### III. Concerns of AI-Guided Communication

17. Have you noticed any biases, such as cultural, emotional, or stylistic, in the suggestions provided by AI? Could you give an example?
18. Do you feel that AI tends to agree with your views or reinforce your existing opinions? If so, how might that influence your communication behavior or thought processes?
19. What do you think are the possible impacts of such biases in AI suggestions on yourself, your relationships, or even on decision-making?
20. Do you have any concerns about using AI in interpersonal communication?